\documentstyle[aps,prb,epsfig]{revtex}
\begin{document}
\twocolumn
\wideabs
{
\title{Infrared Spectroscopy of Quantum Crossbars}
\author{I. Kuzmenko$^{1}$, S. Gredeskul$^{1}$, K. Kikoin$^{1}$, Y. 
Avishai$^{1,2}$}
\address{$^{1}$Department of Physics,\\
$^{2}$The Ilse Katz Center for Mezo and Nanoscale Science and 
Technology,\\
Ben-Gurion University of the Negev, Beer-Sheva }
\date{\today}
\maketitle
\begin{abstract}
Infrared (IR) spectroscopy can be used as an important and effective
tool for probing periodic networks of quantum wires or nanotubes
(quantum crossbars, QCB) at finite frequencies far from the Luttinger
liquid fixed point.  Plasmon excitations in QCB may be involved in
resonance diffraction of incident electromagnetic waves and in optical
absorption in the IR part of the spectrum.  Direct absorption of
external electric field in QCB strongly depends on the direction of
the wave vector ${\bf q}.$ This results in two types of $1D\to 2D$
dimensional crossover with varying angle of an incident wave or its
frequency.  In the case of QCB interacting with semiconductor
substrate, capacitive contact between them does not destroy the
Luttinger liquid character of the long wave QCB excitations.  However,
the dielectric losses on a substrate surface are significantly changed
due to appearance of additional Landau damping.  The latter is
initiated by diffraction processes on QCB superlattice and manifests
itself as strong but narrow absorption peaks lying below the 
damping region of an isolated substrate.
\end{abstract}
} 
\vspace{\baselineskip}
\vspace{\baselineskip}
\section{Introduction}\label{sec:Intro}
Recent achievements in material science and technology have led to
creation of an unprecedented variety of artificial structures that
possess properties never encountered in "natural" quantum objects. 
One of the most exciting developments in this field is fabrication of
two-dimensional (2D) networks by means of self-assembling, etching,
lithography and imprinting techniques \cite{Diehl,Wei}.  Another
development is the construction of 2D molecular electronic circuits
\cite{Luo} where the network is formed by chemically assembled
molecular chains.  Such networks have the geometry of crossbars, and
bistable conformations of molecular chain may be used as logical
elements \cite{Tseng}.  Especially remarkable is a recent experimental
proposal to fabricate 2D periodic grids from single-wall carbon
nanotubes (SWCNT) suspended above a dielectric substrate
\cite{Rueckes}.  The possibility of excitation of a SWCNT by external
electric field together with its mechanical flexibility makes such a
grid formed by nanotubes a good candidate for an element of random
access memory for molecular computing.\\

From a theoretical point of view, such double 2D grid i.e. two
superimposed crossing arrays of parallel conducting quantum wires
\cite{Avron,Avishai,Guinea1} or nanotubes \cite{Mukho1}, represent a
unique nano-object - quantum crossbars (QCB).  Its spectral properties
can not be treated in terms of purely 1D or 2D electron liquid theory. 
A constituent element of QCB (quantum wire or nanotube) possesses the
Luttinger liquid (LL) like spectrum \cite{Bockrath,Egger}.  A single
array of parallel quantum wires is still a LL-like system qualified as
a sliding phase \cite{Mukho1} provided only the electrostatic
interaction between adjacent wires is taken into account.  If an
inter-wire tunneling is possible, the electronic spectrum of an array
is that of 2D Fermi liquid (FL) \cite{Wen1,Schultz1}.\\

Similar low-energy, long-wave properties are characteristic of QCB as
well.  Its phase diagram inherits some properties of a sliding phases
in case when the wires and arrays are coupled only by capacitive
interaction \cite{Mukho1,KGKA1}.  When inter-array electron tunneling
is possible, say, in crosses, dimensional crossover from LL to 2D FL
occurs \cite{Guinea2,Mukho1,Mukho2}.  If tunneling is suppressed and
the two arrays are coupled only by electrostatic interaction in the
crosses, the system possesses the LL zero energy fixed point, and a
rich Bose-type excitation spectrum (plasmon modes) arises at finite
energies in 2D Brillouin zone (BZ)\cite{KGKA1,GKKA}.  These QCB
plasmons can be treated as a set of dipoles distributed within QCB
constituents.  In a single wire the density of the dipole momentum is
proportional to the LL boson field $\theta(x)$ ($x$ is the coordinate
along the wire).\\

Two sets of coupled $1D$ dipoles form unique system which posesses the
properties of $1D$ and $2D$ liquid depending on the type of
experimental probe.  Some possibilities of observation of $1D\to 2D$
crossover in transport measurements were discussed in
Ref.\onlinecite{Mukho1}.  Here we consider various possibilities of
direct observation of plasmon spectra at finite frequencies and wave
vectors by the methods of IR spectroscopy.\\

In transport measurements, the geometrical factors regulate the
crossover from anisotropic to isotropic resistivity of QCB: one may
study the dc response for a field applied either parallel to one of
the constituent arrays or in arbitrary direction.  One may also study
spatially nonuniform response by means of two probes inserted at
different points of QCB and regulate the length scale i.e. the
distance between two probes in comparison with the periods of the
crossbar superlattice.  These methods give information about nearest
vicinity of LL fixed point at $(q,\omega,T)\to 0.$\\

Unlike transport measurements, the methods of infrared spectroscopy
provide an effective tool for investigating the excitation spectrum in
wide enough $(q,\omega)$ area well beyond the sliding phase region. 
We will show that the IR spectroscopy allows scanning of $2D$
Brillouin zone in various directions and thereby elucidates
dimensional crossover in the high symmetry points of the BZ.\\

Several crossover effects such as appearance of non-zero transverse
space correlators and periodic energy transfer between arrays ("Rabi
oscillations") were discussed in our earlier
publications\cite{KKGA,KGKA2}.  The direct manifestation of
dimensional crossover is the response to an external ac
electromagnetic field.  To estimate this response one should note that
the two main parameters characterising the plasmon spectrum in QCB are
the Fermi velocity $v$ of electrons in a wire and the QCB period $a$
(we assume both periods to be equal).  These parameters define both
the typical QCB plasmon wave numbers $q=|{\bf q}|\sim Q=2\pi/a$ and
the typical plasmon frequencies $\omega\sim \omega_{Q}=vQ$.  Choosing
according to Refs.[\onlinecite{Egger,Rueckes}] $v\approx 0.8\cdot
10^{6}$~m/sec and $a\approx 20$~nm, one finds that characteristic
plasmon frequencies lie in the far infrared region $\omega\sim
10^{14}$~sec$^{-1}$, while characteristic wave vectors are estimated
as $q\sim 10^{6}cm^{-1}.$\\

In this paper we study high frequency properties of the simplest
double square QCB (generalization to more complicated geometries is
straightforward).  We start from QCB interacting with an external
infrared radiation.  The plasmon velocity $v$ is much smaller than the
light velocity $c$ and the light wave vector $k$ is three orders of
magnitude smaller than the characteristic plasmon wave vector $Q$
corresponding to the same frequency.  Therefore, infrared radiaton
incident directly on an {\em isolated} array, can not excite plasmons
at all (it could excite plasmon with $\omega\neq 0$).  However in QCB
geometry, each array serves as a diffraction lattice for its partner,
giving rise to Umklapp processes of wave vectors $nQ,$ $n$ integer. 
As a result, excitation of plasmons in the BZ center $q=0$ with
frequencies $nvQ$ occurs.\\

To excite QCB plasmons with $q\neq 0$ one may use an additional
diffraction lattice (DL) with period $A>a$ coplanar to the QCB. Here
the diffraction field contains space harmonics with wave vectors $2\pi
M/A,$ $M$ integer, that enables one to scan plasmon spectrum within
the BZ. Dimensional crossover manifests itself in the appearance of
additional absorption lines when the wave vector of diffraction field
is oriented along specific directions.  In the general case one
observes the single absorption lines forming two sets of equidistant
series.  Instead of that, in the main resonance direction (QCB
diagonal) an equidistant series of split doublets can be observed.  In
the case of higher resonance direction, absorption lines form an
alternating series of singlets and split doublets demonstrating new
type of dimensional crossover related to the frequency change with
direction fixed.\\

We study also QCB interacting with a semiconducting substrate.  This
interaction may be rather strong because surface plasmon waves exist
in the same frequency and wave vector area as plasmons in QCB. This
problem is interesting from at least two points of view.  First,
technological reasons may lead to creation of QCB embedded into thin
layer of semiconductor medium.  Second, the question of stability of
LL regime in QCB against interaction with $2D$ plasmons in substrate,
is itself of great interest.  It will be shown that in spite of the
long range character of 2D screening, the substrate plasmons only
renormalize the velocity of the QCB plasmons and do not destroy LL
character of the QCB spectrum.  At the same time, QCB-substrate
resonance can be observed experimentally in special directions. 
Moreover, at a special frequency, a triple resonance involving both
QCB modes and substrate plasmons is possible.\\

The structure of the paper is as follows.  In section \ref{sec:QCB},
we briefly describe double square QCB and introduce the necessary
definitions.  Interaction of QCB with external field is studied in
section \ref{sec:Absorption}.  In its first part we consider the case
when the incident infrared radiation falls directly on the QCB
(subsection \ref{subsec:Direct}).  In the second part
\ref{subsec:Scan} we study possible scanning of QCB spectrum with the
help of an external DL. Dielectric properties of a combined system
QCB-substrate are studied in the last section \ref{sec:Substrate}.  In
the Conclusion we summarize the results obtained.\\
\section{Double Square QCB}\label{sec:QCB}
\begin{figure}[htb]
\centering
\includegraphics[width=60mm,height=20mm,angle=0,]{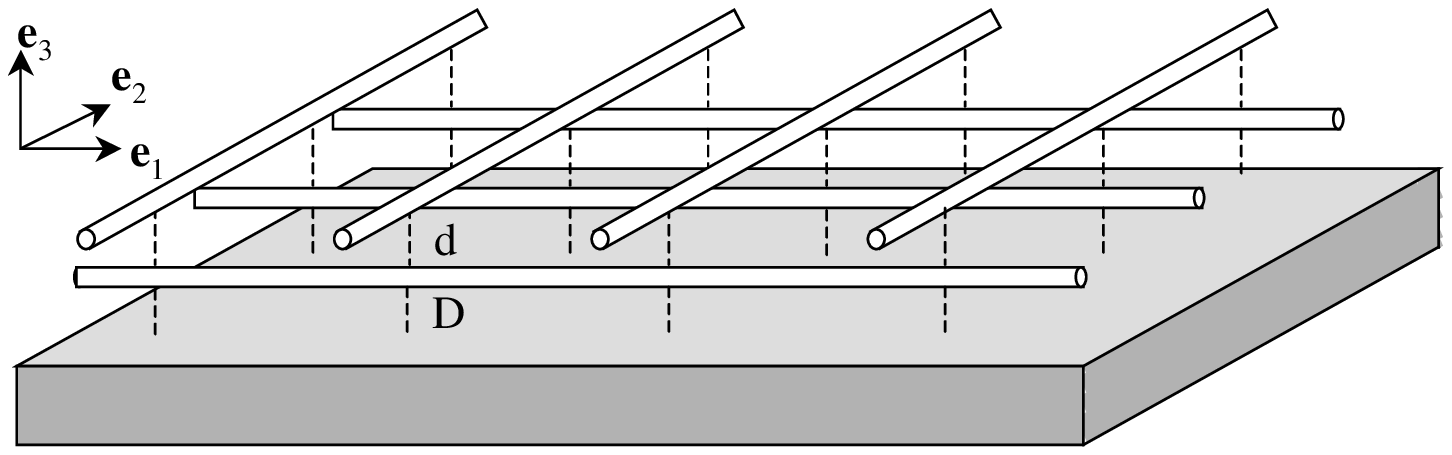} 
\epsfxsize=70mm \caption{QCB on a substrate.  ${\bf e_{j}}$ - basic 
vectors of the coordinate system.  Vector ${\bf e_{1}}$ (${\bf 
e_{2}}$) is oriented along the first (second) array. The inter-array 
distance is $d$ and the distance between substrate and the first 
(lower) array is $D.$}
\label{Substrate}
\end{figure}

A square QCB is a $2D$ grid, formed by two periodically crossed
perpendicular arrays of $1D$ quantum wires or carbon nanotubes. 
Arrays are labeled by indices $j=1,2$ and wires within the first
(second) array are labeled by an integer index $n_{2}$ ($n_{1}$).  In
experimentally realizable setups, QCB is a cross-structure of
suspended single-wall carbon nanotubes lying in two parallel planes
separated by an inter-plane distance $d,$ placed on a dielectric or
semiconducting substrate (see Fig.\ref{Substrate}).  Nevertheless,
some generic properties of QCB may be described under the assumption
that QCB is a genuine $2D$ system.  We choose coordinate system so
that 1) the axes $x_{j}$ and corresponding basic unit vectors ${\bf
e}_{j}$ are oriented along the $j$-th array; 2) the $x_{3}$ axis is
perpendicular to QCB plane; 3) $x_{3}$ coordinate is zero for the
second array, $-d$ for the first one, and $-(d+D)$ for the substrate. 
The basic vectors of the reciprocal superlattice for a square QCB are
$Q{\bf e}_{1,2},$ $Q=2\pi/a$ so that an arbitrary reciprocal
superlattice vector ${\bf m}$ is a sum ${\bf m}={\bf m}_{1}+ {\bf
m}_{2},$ where ${\bf m}_{j}=m_{j}Q {\bf e}_{j},$ ($m_{j}$ integer). 
The first BZ is a square $|q_{1,2}|\leq Q/2.$ \\

A single wire is characterized by its radius $r_{0}$, length $L$, and
LL interaction parameter $g$.  The minimal nanotube radius is
$r_{0}\approx 0.35$~nm\cite{Louie}, maximal nanotube length is
$L\approx 1$~mm, and the LL parameter is estimated as $g\approx 0.3$
\cite{Egger}.  In typical experimental setup\cite{Rueckes} the
characteristic lengths mentioned above have the following values
\begin{eqnarray*}
       d\approx 2 ~{\rm nm},\phantom{aa}L\approx 0.1 ~{\rm mm},
    \label{lengths}
\end{eqnarray*}
so that the inequalities
\begin{eqnarray*}
    r_{0}\ll d\ll a\ll L
    \label{ineq}
\end{eqnarray*}
are satisfied.

The Hamiltonian of QCB interacting with an external field is
\begin{equation}
    H=H_{QCB}+H_{E}.
    \label{Total}
\end{equation}
The QCB interaction $H_{E}$ with an external
electric field ${\bf E}=(E_{1},E_{2},E_{3})$) is nothing but an
energy of a set of QCB dipoles in this field
\begin{eqnarray*}
    H_{E}&=&-e\sqrt{2}
    \left\{
    \sum_{n_{2}}\int
dx_{1}E_{1}(x_{1},n_{2}a,-d)\theta_{1}(x_{1},n_{2}a,-d)+\right.\nonumber\\
    &&\left.\sum_{n_{1}}\int dx_{2}E_{2}(n_{1}a,
x_{2},0)\theta_{2}(n_{1}a,x_{2},0),
    \right\}
    \label{eq:DipInt}
\end{eqnarray*}
where $\theta_{j}$ is one of the two conventional canonically conjugate
boson fields $\pi_{j},$ $\theta_{j}.$\\

The QCB Hamiltonian
\begin{equation}
    H_{QCB}=H_{1}+H_{2}+H_{12}.
    \label{HamiltTot}
\end{equation}
consists of three terms.  The first of them describes LL in the first
array
\begin{eqnarray*}
{H}_{1} & = &
         \frac{\hbar v}{2}\sum_{{n}_{2}}
              \int\limits_{-L/2}^{L/2} {dx}_{1}
              \biggl\{
                   g{\pi}_{1}^{2} \left( x_1,
                   {n}_{2}a,-d\right)
                   \\
               &+ &    \frac{1}{g}
                   \left(
                        {\partial}_{{x}_{1}}
                        {\theta}_{1}
                        \left(
                             {x}_{1},{n}_{2}a,-d
                        \right)
                   \right)^2
         \biggr\}.
\end{eqnarray*}
The Hamiltonian $H_{2}$ is obtained from $H_{1}$ after permutation
$1\leftrightarrow 2$ and replacement $-d\rightarrow 0$ in the
arguments of the fields.\\

The inter-array interaction is described by the last term in 
Eq.(\ref{HamiltTot})
\begin{eqnarray}
   &&H_{12}  =  V_{0}\sum\limits_{{n}_{1},{n}_{2}}
            \int dx_1 dx_2
            \zeta\left(
                 \frac{x_1-n_1a}{r_{0}}
                 \right)
            \zeta\left(
                 \frac{n_2a-x_2}{r_{0}}
                 \right)\nonumber \\
		 &&\times \partial_{x_1}\theta_1(x_1,n_2a,-d)
		 \partial_{x_2}\theta_2(n_1a,x_2,0),
		 \phantom{aa}V_{0}=\frac{2e^{2}}{d}.
     \label{Interaction}
\end{eqnarray}
It results from a short-range contact capacitive coupling in the
crosses of the bars.  The dimensionless envelope function (introduced
phenomenologically) $\zeta(\tau_{j})$ describes re-distribution of a
charge in a tube $j$ induced by the interaction with tube $i.$ This
function is of order unity for $|\tau|\sim 1$ and vanishes outside
this region so that the dimensionless integral
\begin{eqnarray*}
    \int\zeta(\tau)e^{ikr_{0}\tau}d\tau\sim 1
    \label{barzeta}
\end{eqnarray*}
is of order unity for all $|k|$ smaller than a certain ultraviolet
cutoff.  For simplicity, we put it equal to unity in what follows.\\

Before turning to investigation of QCB interaction with an external
field, we briefly describe the spectral properties of the QCB
itself\cite{KGKA2}.  The QCB Hamiltonian (\ref{HamiltTot}) is a
quadratic form in terms of the field operators, so it can be
diagonalized exactly.  Such a procedure is rather cumbersome. 
However, due to the separability of the interaction
(\ref{Interaction}) the spectrum can be described analytically. 
Moreover, being small, this interaction grossly conserves the
unperturbed $1D$ systematics of levels and states, at least in the low
energy region corresponding to the first few energy bands.  This means
that perturbed eigenstates could be described in terms of the same
quantum numbers (array number, band number and quasimomentum) as the
unperturbed eigenstates of an ``empty'' lattice.  Such a description
fails in two specific regions of a reciprocal space ${\bf k}=k_{1}{\bf
e}_{1}+k_{2}{\bf e}_{2}$.  The first of them is the vicinity of high
symmetry lines $k_{j}=nQ/2$ with $n$ integer (the lines with $n=\pm 1$
include BZ boundaries).  Around these lines, the {\em interband}
mixing is significant and two modes from the adjacent bands are
degenerate.  The second region is the vicinity of the resonant lines
$k_{1}\pm k_{2}=nQ$ where the eigenfrequencies of unperturbed plasmons
from the same band $s$ propagating along two arrays coincide
$\omega_{1s}(q_{1})=\omega_{2s}(q_{2}).$ Around resonant lines {\em
inter-array} mixing is significant and at these lines two modes
corresponding to different arrays are degenerate.\\

Inter-array interaction introduces real two dimensionality into the
problem, lifts the degeneracy of bare modes and splits degenerate
frequencies.  Therefore the simplest way to probe 2D nature of QCB is
to observe this splitting in the corresponding spectral region. 
Possible ways of such an observation are discussed in the next two
sections.\\
\section{Infrared Light Absorption by QCB}\label{sec:Absorption}
\subsection{Long Wave Absorption}\label{subsec:Direct}
\begin{figure}[htb]
\centering
\includegraphics[width=40mm,height=60mm,angle=0,]{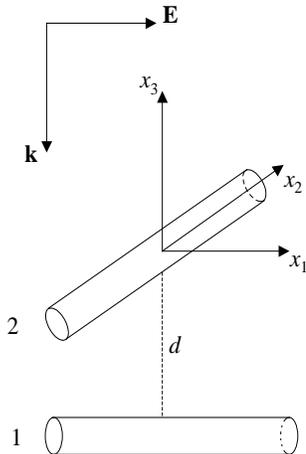}
\epsfxsize=70mm \caption{The incident field orientation with respect
to QCB. The axes $x_{1}$ and $x_{2}$ are directed along the
corresponding arrays, and $d$ is the inter-array vertical distance
(along the $x_{3}$ axis).}
\label{Cross1}
\end{figure}

In the case of a dielectric substrate transparent in the infrared
region, one can treat QCB as an isolated grid (without substrate)
interacting directly with the incident radiation.  Consider the simplest
geometry (see Fig.\ref{Cross1} for details) where an external wave
falls normally onto QCB plane, and its electrical field
$${\bf E}=E_{0}{\bf e}_{1}\cos{({\bf k}{\bf r}-\omega t)}$$
is parallel to the lower (first) array.  In this geometry the field
${\bf E}$ is {\it longitudinal} for array $1$ and {\it transverse} for
the array $2$.  The eigenfrequencies of transverse modes in array $2$
substantially exceed the IR frequency of the incident wave and even
the standard LL ultraviolet cutoff frequency.  Thus, the incident wave
can be treated as a static polarization field for this array, and the
factor $\cos{\omega t}$ can be omitted.  Then, the polarization waves
in array $2$ form a longitudinal diffraction field for array $1$ with
quasi wave vectors $nQ$ ($n$ integer).  Further, the characteristic
order of magnitude $Q$ of a QCB plasmon wave vector is much larger
than the wave vector ${\bf k}$ of the incident light, and we put the
latter equal to zero from the very beginning.  Then, the light
wavelength is much larger than a nanotube diameter and the geometrical
shadow effect can be neglected.  As a result the total field which
affects array 1 consists of an external field and a diffraction field
produced by a static charge induced in array 2.\\

To calculate diffraction field consider first the field ${\bf E}^{0}$
produced by the quantum wire of array 2 which is located at
$x_{1}=x_{3}=0$ and labeled by $n_{1}=0$.  The large distance between
the wire under consideration and its neighbour partners from the same
array allows us to neglect the influence of the charges induced on
them.  The static potential on the surface of the wire includes
external potential of an incident field and the potential $\Phi^{0}$
of the charge induced on the wire.  On the other hand this static
potential should be equal to a constant which we choose to be zero. 
In cylindrical coordinates $r,\vartheta, x_{2},$
$x_{1}=r\cos\vartheta, x_{3}=r\sin\vartheta,$ this condition reads
\begin{eqnarray}
    \Phi^{0}(R_0,\vartheta,x_{2}) = E_0r_0\cos\vartheta.
    \label{Pot}
\end{eqnarray}
Outside the wire, induced potential $\Phi^{0}$ satisfies Laplace
equation $\Delta\Phi^{0}=0.$ Solving this equation with boundary
condition (\ref{Pot}) we obtain the static part of the induced
potential
\begin{eqnarray*}
    \Phi^{0}(r,\vartheta,x_{2})=\frac{E_0r_0^2}{r}\cos\vartheta
    \label{IndPhi}
\end{eqnarray*}
and the corresponding static part of the induced field along
$x_{1}$ direction
\begin{eqnarray*}
    E_{1}^{0}(x_{1},x_{3})=-E_{0}
    \frac{r_0^2\left(x_{3}^2-x_{1}^2\right)}
           {\left(x_{3}^2+x_{1}^2\right)^2}.
    \label{IndField}
\end{eqnarray*}
The first component of the difraction field is the sum of the fields
induced by all wires of the upper array,
\begin{eqnarray}
        E_{1}(x_{1};t)&=&\cos\omega t\sum_{n_{1}}
        E_{1}^{0}(x_{1}-n_{1}a,-d)\nonumber\\
    &=&-E_{0}\cos\omega t
    \sum_{n_{1}}
    \frac{r_0^2\left(d^2-(x_{1}-n_{1}a)^2\right)}
           {\left(d^2+(x_{1}-n_{1}a)^2\right)^2}.
    \label{Field}
\end{eqnarray}
This field is a periodic function of $x_{1}$ with period $a.$
Therefore, its Fourier expansion contains only wave vectors
$k_{1n}=nQ$ ($n$ is the order of diffraction).  This means that only
frequencies $\omega_{n}=nvQ$ can be excited.  In this case it is more
convenient to expand the field over Bloch eigenfunctions of an
``empty'' wire\cite{KGKA2}.  These functions are labeled by
quasimomentum $q_{1},$ $|q_{1}|\leq Q/2,$ and the band number $s$. 
The expansion includes only $q_{1}=0$ components and has the form
\begin{eqnarray*}
    E_{1}(x_{1};t)=\cos\omega t\sum_{s}
    E_{[s/2]}
    u_{s}(x_{1}),
    \label{Expansion1}
\end{eqnarray*}
where
 \begin{eqnarray*}
    u_{s}(x)=
    \exp\left(
    iQx[s/2](-1)^{s-1}
    \right)
    \label{Bloch}
 \end{eqnarray*}
is the $q_{1}=0^{+}$ Bloch amplitude $u_{sq_{1}}(x)$ within the $s$-th
band, $[\ldots]$ is the entire part symbol, and
\begin{equation}
    E_{n}=-E_{0}
    \frac{\pi r_{0}^{2}}{ad}
    nQde^{-nQd}.
    \label{Expansion2}
\end{equation}
The excited eigenfrequency $\omega_{n}=\omega_{[s/2]}$ belongs
simultaneously to the top of the lower even band with number $s=2n$
and to the bottom of the upper odd band with number $s=2n+1$ (this is
the result of ${\bf{E}}(x)$ parity).  Incident field corresponds to
$n=0$ and we do not take it into account.\\

Turning to ${\bf q},s$ representation with the help of the expansion 
\begin{eqnarray}
 \theta_1(x_1,n_2a)=\frac{\sqrt{a}}{L}
 \sum\limits_{s{\bf{q}}}\theta_{1s{\bf{q}}}
 e^{i(q_1x_1+q_2n_2a)}\nonumber\\
 \times\exp\left(
    iQx_1[s/2](-1)^{s-1}{\rm sign}q_1
    \right),
    \label {Fourier}
\end{eqnarray}
and similarly for $\theta_{2}$ and $\pi_{1,2},$ one easily sees
that only the ${\bf q}={\bf 0}$ components are involved in interaction
with the incident radiation.  The corresponding part of the
Hamiltonian (\ref{Total}) has the form ( zero quasimomentum index is
omitted):
\begin{eqnarray*}
 H' & = & \frac{\hbar}{2}\sum_{j,s}
               \left(
                    {v}{g}
                    \pi_{js}^{\dagger}
                    \pi_{js}
                    +\frac{\omega^{2}_{[s/2]}}{vg}
                    \theta_{js}^{\dagger}
                    \theta_{js}
               \right)
               \nonumber\\
        && +\frac{\hbar \phi}{vg}\sum_{ss'}\left(-1\right)^{s+s'}
            \omega_{[s/2]}
        \omega_{[s'/2]}
            \left(
            \theta_{1s}^{\dagger}\theta_{2s'}+
            \theta_{2s'}^{\dagger}\theta_{1s}
            \right)
            \nonumber\\
       && -\frac{eL}{\sqrt{2a}}\cos\omega t
            \sum_{s}E_{[s/2]}
            \left(
            \theta_{1s}^{\dagger}+\theta_{1s}
            \right),
            \label{Involved}
\end{eqnarray*}
where
\begin{eqnarray*}
    \phi =\frac{gV_{0}r_{0}^{2}}{{\hbar}va},
    \phantom{aaa}\phi\sim0.007
    \label{phi}
\end{eqnarray*}
is the dimensionless inter-array interaction.  Consider the initial
frequency $\omega$ close to $\omega_{n}.$ In a resonant approximation,
only four equations of motion for the ``coordinate'' operators
$\theta_{s}$ with $s=2n,2n+1$ are relevant
\begin{eqnarray}
  \ddot{\theta}_{1,2n}+\omega_n^2{\theta}_{1,2n}+
  \phi\omega_n^2
  \left({\theta}_{2,2n}-{\theta}_{2,2n+1}\right)
  &=& Lf_n\cos\omega t,
  \nonumber\\
  \ddot{\theta}_{1,2n+1}+\omega_n^2{\theta}_{1,2n+1}-
  \phi\omega_n^2
  \left({\theta}_{2,2n}-{\theta}_{2,2n+1}\right)
  &=& Lf_{n}\cos\omega t,
  \nonumber\\
  \ddot{\theta}_{2,2n}+\omega_n^2{\theta}_{2,2n}+
  \phi\omega_n^2
  \left({\theta}_{1,2n}-{\theta}_{1,2n+1}\right)
  &=& 0,
  \nonumber\\
  \ddot{\theta}_{2,2n+1}+\omega_n^2{\theta}_{2,2n+1}
  -\phi\omega_n^2
  \left({\theta}_{1,2n}-{\theta}_{1,2n+1}\right)
  &=& 0,
  \label{EqMot1}
\end{eqnarray}
where $$f_n=\frac{\sqrt{2}{v}g{e}}{\hbar\sqrt{a}}E_{n}.$$
The homogeneous part of this system defines four eigenfrequencies
\begin{eqnarray*}
    \omega_{gg}& =& \omega_{ug}=\omega_{n},\nonumber\\
    \omega_{uu}& \approx & \omega_{n}(1-\phi),\nonumber\\
    \omega_{gu}& \approx& \omega_{n}(1+\phi).
    \label{Frequencies}
\end{eqnarray*}
The corresponding eigenvectors are symmetrized combinations of the
four operators which enter Eq.(\ref{EqMot1}).  They have a fixed
parity with respect to permutation of arrays (the first index) and neighboring bands
(the second index).  Only two modes (even with respect to
band index)
\begin{eqnarray*}
  \theta_{gg}=\frac{1}{2}
  \left({\theta}_{1,2n}+
        {\theta}_{1,2n+1}+
        {\theta}_{2,2n}+
        {\theta}_{2,2n+1}
  \right),\\
  \theta_{ug}=\frac{1}{2}
  \left({\theta}_{1,2n}+
        {\theta}_{1,2n+1}-
        {\theta}_{2,2n}-
        {\theta}_{2,2n+1}
  \right)
\end{eqnarray*}
interact with an external field.  Therefore only the unperturbed
frequency $\omega_{n}=\omega_{gg}=\omega_{ug}$ will be absorbed.  The
two equations of motion for the operators $\theta_{gg,ug}$ have the
same form
\begin{eqnarray*}
  \ddot{\theta}_{\alpha}
  +2\gamma\dot{\theta}_{\alpha}
  +\omega_{n}^2{\theta}_{\alpha}
  = Lf_{n}\cos\omega t,
  \label{EqMotDiag}
\end{eqnarray*}
where $\alpha=gg,ug.$ Employing standard procedure in the vicinity of
the resonance $|\omega-\omega_n|\ll\omega_n$ immediately yields the
relative absorption of Lorentz type
\begin{eqnarray}
  \frac{\Delta I_{n}}{I_{0}}&=&
           2g\frac{e^{2}}{\hbar c}
           \left(\frac{{\pi}r_0^2}{{a}{d}}\right)^2\nonumber\\
           &&\times\frac{\gamma vQ}{\left(\omega-\omega_n\right)^2+\gamma^2}
           \left[
       nQde^{-nQd}
       \right]^{2},
  \label{EnerAbsorp}
\end{eqnarray}
where
$$
 I_0=\frac{cL^2}{4\pi}E_0^2
$$
is the energy of light that falls on the QCB per unit time.\\

Due to the exponential term in the r.h.s of Eq.(\ref{Expansion2}),
$E_{n}$ decreases fast with $n$ and only the first few terms
contribute to absorption.  The characteristic dimensionless scale of
the induced field ${{r}_0^2}/{(ad)}$ for typical values of QCB
parameters equals $0.004$.  We tabulate below the lowest dimensionless
Fourier components of the induced field.
\begin{center}
\begin{tabular}{| c | c | c | c | c | c |}
 \hline
 n & 1 & 2 & 3 & 4 & 5 \\ \hline
 $-\displaystyle{\frac{ad}{r_0^2}\frac{{E}_{n}}{E_0}}$
 & 1.05306 & 1.12359 & 0.89914 & 0.63957 & 0.42650 \\
 \hline
\end{tabular}
\end{center}
The results show that one can hope to probe at least the first five
spectral lines corresponding to $\omega_{n}$ with $n=1,2,\ldots,5.$ \\

The width of the absorption line (\ref{EnerAbsorp}) is governed by an
attenuation coefficient $\gamma.$ It was introduced phenomenologically
but one can (at least qualitatively) estimate its value.  The
attenuation is caused by decay of plasmon into phonons.  The one
phonon decay of the plasmon with wave number $k$ and frequency
$\omega=v|k|$ into a single phonon with the same $\omega$ and $k$
occurs in a single point in $1D$ and does not yield finite attenuation
at all.  Multiphonon decay is weak because of the small anharmonic
coupling within the wire.  As a result, the form of the absorption
lines should be determined mainly by the instrumental linewidth.\\
\subsection{Scanning of the QCB Spectrum within BZ}\label{subsec:Scan}
\begin{figure}[htb]
\centering
\includegraphics[width=70mm,height=49mm,angle=0,]{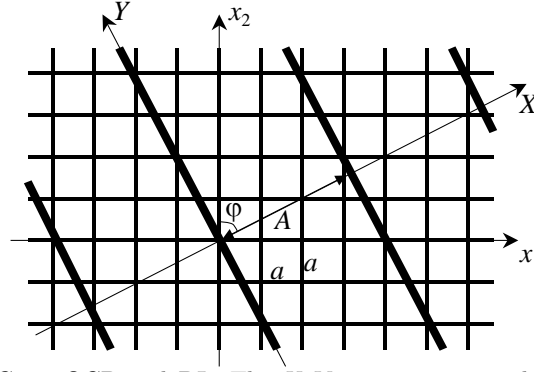} 
\epsfxsize=70mm \caption{QCB and DL. The $X,Y$ axes are oriented along 
the DL stripes and the wave vector ${\bf K}$ of the diffraction field 
respectively. The DL (QCB) period is $A$ ($a$).}
\label{QCB-DLFig}
\end{figure}

Within a geometry considered in the previous subsection, one can probe
plasmon spectrum only at the BZ center.  To study plasmons with
nonzero wave vectors one should add to the system an external
diffraction lattice namely a periodic array of metallic stripes
parallel to the $Y$ axis (see Fig.\ref{QCB-DLFig}).  The DL plane
$Z=0$ is parallel to the QCB planes $Z=-D$ for the upper second array
and $Z=-(D+d)$ for the lower first array (the $Z$ axis is parallel to
the $x_{3}$ axis).  The distance $D$ between DL and second array is of
the same order that the inter-array distance $d=2$~nm.  The angle
between DL wires and the {\em second} array is $\varphi$
($0<\varphi<\pi/2$).  To get a wave number $K$ of a diffraction field
much smaller than $Q$ one needs a DL with a period $A$ much larger
than the QCB period $a$.  In the following numerical estimations we
choose $A\approx200$~nm.\\

Consider an incident field with electric vector ${\bf E}=E_{0}{\bf
e}_{X} \cos({\bf k}{\bf r}-\omega t)$ oriented along the $X$ axis
(perpendicular to the DL wires).  The radius $R_{0}$ of a DL wire is
assumed to be not much larger than the nanotube radius
$r_{0}.$ In this case light scattering on the DL is similar to that 
considered in subsection \ref{subsec:Direct}).  Then the diffraction
field is concentrated along the $X$ direction and has the form (compare
with Eq.(\ref{Field}))
\begin{eqnarray*}
 E_X(X,Z,t) =
 -E_0\cos\omega t\times\nonumber\\
 \sum_{N}
      \frac{R_0^2(Z^2-(X-NA)^2)}
           {(Z^2+(X-NA)^2}.
 \label{FieldDL}
\end{eqnarray*}
The Fourier transform of the diffraction field is
\begin{eqnarray*}
 {E_{X}}({\bf K},Z)=-E_0
      \frac{\pi R_0^2}{A|Z|}
      |KZ|e^{-|KZ|},
 \label{Fourier1}
\end{eqnarray*}
where
$$
{\bf K}(M)=K{\bf e}_{X}=(K_{1},K_{2})=
\frac{2\pi M}{A}(\sin\varphi, \cos\varphi)
$$
with positive integer $M$.  This means that all the points ${\bf K}$
lie on the same ray oriented along the positive direction of the $X$
axis.  The vector ${\bf K}(M)$ for a fixed $M$ can be uniquely
represented as a sum of quasimomenta lying in the first BZ and two
reciprocal lattice vectors
\begin{eqnarray*}
    {\bf K}(M)={\bf q}(M)+{\bf m}_{1}(M)+{\bf m}_{2}(M).
    \label{direction}
\end{eqnarray*}
The field components
\begin{eqnarray}
 E_{1{\bf{K}}} &=&{E_{X}}({\bf K},D+d)\sin\varphi,\nonumber\\
 E_{2{\bf{K}}} &=&{E_{X}}({\bf K},D)\cos\varphi
 \label{Components}
\end{eqnarray}
parallel to the quantum wires, can excite plasmons and contribute to
the absorption process.\\

The Hamiltonian (\ref{Total}) of QCB
interacting with an external field in Fourier
\begin{eqnarray}
      H_j &=&
      \frac{\hbar}{2}\sum_{{\bf{qm}_j}}
      \bigg\{
           {v}{g}\pi_{j{\bf q}+{\bf m}_j}^{\dag}
           \pi_{j{\bf q}+{\bf m}_j}+\nonumber\\
      &&+
           \frac{\omega^2_{q_j+m_jQ}}{{v}{g}}
           \theta_{j,{\bf q}+{\bf m}_j}^{\dag}
           \theta_{j,{\bf q}+{\bf m}_j}
      \bigg\},\ \ \ j=1,2,
      \label{arrays}
	  \\
      H_{12} &=& \frac{\hbar}{{v}{g}}\sum_{{\bf m},{\bf q}}
             \Phi_{{\bf q}+{\bf m}}
             \theta_{1,{\bf q}+{\bf m}_1}^{\dag}
             \theta_{2,{\bf q}+{\bf m}_2},
             \label{inter}\\
      H_{E}&=& \frac{\hbar{L}}{2vg}
      \sum_{{\bf K}}\Big[
      f_{1,{\bf{K}}}\left(
      \theta_{1,{\bf K}}+
      \theta_{1,{\bf K}}^{\dag}
      \right)+
      \nonumber\\&&+
      f_{2,{\bf{K}}}\left(
      \theta_{2,{\bf K}}+\theta_{2,{\bf K}}^{\dag}
      \right)\nonumber 
      \Big],
      \label{Hams}
\end{eqnarray}
where
\begin{eqnarray}
      \phantom{aaa}
      \Phi_{{\bf{k}}}&=&\phi\xi_{k_1}\xi_{k_2},\phantom{a}
      \xi_{k}=\omega_{k}{\rm sign}k,\phantom{a}
      \omega_{k}=v|k|,\phantom{aaa}\nonumber\\
      f_{j,{\bf{K}}}&=&\frac{\sqrt{2}vge}{\hbar\sqrt{a}}E_{j,{\bf{K}}},
      \phantom{a}{\bf{m}}={\bf{m}}_1+{\bf{m}}_2.
\label{Explan}
\end{eqnarray}
In this subsection we are interested not in the form of the absorption
line but only in the resonant frequencies.  Therefore, we did not introduce
any phenomenological attenuation.  The equations of motion
for boson fields have the form
\begin{eqnarray}
  &&\ddot{\theta}_{1,{\bf{q}}+{\bf{m}}_1}+
  \omega_{q_1+m_1Q}^2{\theta}_{1,{\bf{q}}+{\bf{m}}_1}+
  \nonumber\\+
  \sum_{m_2}&&\Phi_{{\bf{q}}+{\bf{m}}}
  {\theta}_{2,{\bf{q}}+{\bf{m}}_2}
  = L\sum_{M,m_2}f_{1,{\bf{K}}}\delta_{{\bf{K}},{\bf{q}}+{\bf{m}}},
  \nonumber\\
  &&\ddot{\theta}_{2,{\bf{q}}+{\bf{m}}_2}+
  \omega_{q_2+m_2Q}^2{\theta}_{2,{\bf{q}}+{\bf{m}}_2}+
  \nonumber\\+
  \sum_{m_1}&&\Phi_{{\bf{q}}+{\bf{m}}}
  {\theta}_{1,{\bf{q}}+{\bf{m}}_1}
  = L\sum_{M,m_1}f_{2,{\bf{K}}}\delta_{{\bf{K}},{\bf{q}}+{\bf{m}}}.
  \label{EqMot2}
\end{eqnarray}
Only the first few terms in the sum over $K$ in the r.h.s. of Eq. 
(\ref{EqMot2}) really excite the QCB plasmons.  Indeed, the diffraction
field (\ref{Fourier}) is proportional to the same dimensionless
function of the type $te^{-t}$ ($t=|KZ_{j}|$) as in the previous
subsection (see Eq.(\ref{Expansion2})).  This function has its maximum
at $t=1$ and differs significantly from zero for $0.2<t<2.7$.  For
$a=20$~nm, $D=2$~nm, it is of order unity within the interval
$0.18Q<|K|<2.13Q$ for the first array ($Z_{1}=D+d$), and within the
interval $0.36Q<|K|<4.26Q$ for the second array ($Z_{1}=D$).  This
means that one can excite the modes of four lower bands ($K<2Q$) of
the first array and the modes of eighth lower bands ($K<4Q$) of the
second array.\\

According to Eqs.  (\ref{Components}) the field $E_{j{\bf K}(M)}$ is
coupled with plasmons of wave vectors ${\bf q}+{\bf m}_{j}={\bf
q}(M)+{\bf m}_{j}(M)$ within the $j$-th array.  The nature of the
excited plasmons as well as their frequencies depend on the direction
of the vector ${\bf K}(M).$ For simplicity we restrict ourselves by
acute angles $0<\varphi<\pi/2$ describing orientation of both DL and
vector ${\bf K}(M)$.  There are four kinds of dimensional crossover
depending on specific directions in the BZ. Each type of crossover is
characterized by its own set of absorption lines.  The first one takes
place in a common case when ${\bf K}(M)$ for any $M$ never reaches
neither a resonant direction nor the BZ boundary.  The second case
corresponds to the bisectorial direction $\varphi=\pi/4$ where the
main resonant condition $\omega(K_{1})=\omega(K_{2})$ is fulfilled. 
The third group of directions is determined by another resonant
condition $\omega(K_{1})=\omega(nQ\mp K_{2}).$ Finally, the fourth
group is formed by directions intersecting with the BZ boundaries for
some values of $M.$ In what follows we consider these four cases
separately.\\

{\bf 1.} In the general case, the points ${\bf K}(M)$ for all $M$ are
far from the BZ diagonals and boundaries.  Therefore each of them
corresponds to two plasmons mostly propagating along the $j$-th array,
$j=1,2,$ with unperturbed frequencies $\omega_{K_{j}(M)}=vK_{j}(M)$. 
The inter-array interaction slightly renormalizes the eigenfrequencies
\begin{eqnarray*}
 \omega_{1{\bf{K}}}^2 &=& \omega_{K_1}^2+
            \phi^2\sum_{m_2}
            \frac{\omega_{K_1}^{2}
                  \omega_{K_2+m_2Q}^{2}}
                 {\omega_{K_1}^2-\omega_{K_2+m_2Q}^2},\\
 \omega_{2{\bf{K}}}^2 &=& \omega_{K_2}^2+
            \phi^2\sum_{m_1}
            \frac{\omega_{K_2}^{2}
                  \omega_{K_1+m_1Q}^{2}}
                 {\omega_{K_2}^2-\omega_{K_1+m_1Q}^2}.
\end{eqnarray*}
Thus, increasing the frequency of an incident light one observes a set
of single absorption lines that consists of two almost equidistant
subsets with frequencies corresponding to excitation of plasmons in the
first or second arrays. The distances between adjacent lines within
each subset are
\begin{eqnarray*}
\Delta\omega_{1}&=& v\Delta K_{1}=2\pi v \sin\varphi/A,\nonumber\\
\Delta\omega_{2}&=& v\Delta K_{2}=2\pi v \cos\varphi/A,
\end{eqnarray*}
and their ratio depends on the DL orientation $\varphi$ only
$$\frac{\Delta\omega_{1}}{\Delta\omega_{2}}=\tan\varphi.$$
\\

{\bf 2.} In the resonant case $\varphi=\pi/4,$ the relation
$K_1(M)=K_2(M)$ is satisfied for all $M$.  Therefore modes propagating
along the two arrays are always degenerate.  Inter-array interaction
lifts the degeneracy.  Indeed, in the resonant approximation, the
coupled equations of motion for the field operators read
\begin{eqnarray}
  \ddot{\theta}_{1{\bf{K}}}+\omega_{K_1}^2\theta_{1{\bf{K}}}+
  \phi\omega_{K_1}^2\theta_{2{\bf{K}}}
  &=& f_{1{\bf{K}}},
 \nonumber\\
  \ddot{\theta}_{2{\bf{K}}}+\omega_{K_1}^2\theta_{2{\bf{K}}}+
  \phi\omega_{K_1}^2\theta_{1{\bf{K}}}
  &=& f_{2{\bf{K}}}.\nonumber 
  \label{Eq-Mot-Res}
\end{eqnarray}
After symmetrization
$\theta_{g,u}=(\theta_{1{\bf{K}}}\pm\theta_{2{\bf{K}}})/\sqrt{2}$,
they have the same form
$$
  \ddot{\theta}_{\alpha}+\omega_{\alpha}^2\theta_{\alpha}
  =f_{\alpha},
$$
where
$$
\omega_{g,u}=\omega_{K}\left(1\pm\frac{1}{2}\phi\right)
$$
are the renormalized frequencies, $\alpha=g,u$, and
$f_{g,u}=({f}_{1{\bf{K}}}\pm{f}_{2{\bf{K}}})/\sqrt{2}$.  The
amplitudes $f_{g,u}$ are of the same order of magnitude because the
distances $D$ and $d$ are different but have the same order of
magnitude.  As a result, increasing the frequency of an incident light
one observes an equidistant set of absorption doublets with distance
$\pi\sqrt{2}v/A$ between adjacent doublets.\\

{\bf 3.} Consider now the directions $\varphi$ determined by the
equation
\begin{eqnarray*}
    \sin\left(\varphi\pm\frac{\pi}{4}\right)=
\frac{nA}{\sqrt{2}M_{0}a},
    \label{varphi}
\end{eqnarray*}
where $n$ and $M_{0}$ are mutually prime integers.  For this
direction, two components of the first $M_{0}-1$ points ${\bf K}(M)$
do not satisfy any resonant condition while the $M_{0}$-th one does
\begin{equation}
     K_1(M_{0})\pm K_2(M_{0})=nQ.
     \label{resonant}
 \end{equation}
With increasing $M$ this situation is reproduced periodically so that
all points ${\bf K}(pM_{0})$ with $p$ integer satisfy a similar
condition with $pn$ standing instead of $n,$ while all intermediate
points are out of resonance.\\

In the zero approximation with respect to the inter-array interaction
we expect to observe two set of absorption lines with frequencies
$p\omega_{j}=vK_{j}(pM_{0}),$ $j=1,2,$ corresponding to excitation of
plasmons within the $pm_{j}(M_{0})$-th band of the $j$-th array.  The
ratio of the frequencies $\omega_{j}$ is defined by DL orientation
\begin{eqnarray*}
    \frac{\omega_{1}}{\omega_{2}}=\tan\varphi.
    \label{relation}
\end{eqnarray*}
However, due to the resonance condition (\ref{resonant}), plasmon in
the first array with wave vector $K_{1}(pM_{0})$ and frequency
$\omega_{1}=vK_{1}(pM_{0})$ is coupled with plasmon in the second
array with the same frequency and wave vector $K'_{2}=\mp(npQ-
K_{1}(pM_{0}))$ (inter-array degeneracy).  Similarly, plasmon in the
second array with wave vector $K_{2}(pM_{0})$ and frequency
$\omega_{2}=vK_{2}(pM_{0})$ is coupled with plasmon in the first array
with the same frequency and wave vector $K'_{1}=npQ\mp K_{2}(pM_{0}).$
This degeneracy of two modes corresponding to the same band but to
different arrays is lifted by the inter-array interaction.  As a
result one has two sets of doublets instead of two sets of single
lines.\\

Thus, for such orientation of DL, increasing frequency of an incident
wave one should observe two equidistant sets of single absorption lines
with two sets of equidistant doublets built in these series
\begin{eqnarray*}
 \omega_{1{\bf{K}}} &=&
\omega_{K_1(pM_{0})}\left(1\pm\frac{1}{2}\phi\right),\\
 \omega_{2{\bf{K}}} &=&
\omega_{K_2(pM_{0})}\left(1\pm\frac{1}{2}\phi\right).
 \label{omega-res}
\end{eqnarray*}
In the case $n=1$ the lower doublet lies in the first energy band,
whereas the upper one lies in the second band.  For $A/a=10$ (that
corresponds to the realistic values of the parameters $a=20$~nm and
$A=200$~nm) the lowest doublet ($p=1$) will be observed for example
for integer $M_{0}=8,$ at the angle $\varphi(8)\approx 17^{\circ}$,
around frequencies $\omega_{1}(8)=0.76vQ,$
$\omega_{2}(8)=0.24vQ.$\\

{\bf 4.} It seems that a similar behavior will be manifested in
the case when the points ${\bf K}_{pM}$ lie at one of the BZ
boundaries, i.e.  satisfy the relation
\begin{eqnarray*}
    K_{j}(pM_{j})=\frac{npQ}{2}
    \label{boundaries}
\end{eqnarray*}
with some specific values $j,$ $M_{j}$ and $n.$ Such situation is
realized at specific angles that depend on the integers $j,n,M_{j}.$
In the vicinity of the points ${\bf K}(pM_{j})$ two frequencies
corresponding to the unperturbed modes of the $j$-th array from the
$np$-th and $(np+1)$-th bands coincide.  This is the case of
inter-band degeneracy that is also lifted by inter-array interaction. 
Due to the square symmetry (invariance with respect to
${x_j}\to{-x_j}$ inversion), only one of the two components with
frequency $\omega=v|K_{j}(pM_{j})|$ may be excited by a diffraction
field.  Therefore, this case does not differ from the case {\bf 1}
considered above and two sets of equidistant single lines can be
observed.\\

We emphasize that studying absorption of light by QCB one can expose,
beyond the standard\cite{KGKA2} dimensional crossover with respect to
an angle (direction), also occurrence of a new type of crossover with
{\em an external frequency} as a control parameter.  This occurs for
special directions of type ${\bf 3}$ where, with increasing frequency,
the set of single lines is periodically intermitted by doublets.\\
\section{QCB on a Semiconducting Substrate}\label{sec:Substrate}
\subsection{Hamiltonian}\label{subsec:Hamilt}
An alternative method of examination of the plasmon spectrum in QCB is
the study of its interaction with the substrate, whose electrodynamic
properties are well known.  An excellent material for this purpose is
a III-V semiconductor.  Surface plasmons in semiconductor slab can be
excited in a controllable way (see Refs.\cite{Agranovich,Boardman} and
literature therein), and these plasmons can be in resonance with Bose
excitations in QCB.\\  

In this section we consider dielectric properties of QCB placed on a
semiconducting substrate (Fig.\ref{Substrate}).  Any surface wave
excited in the substrate is coupled with QCB-plasmon modes due to the
substrate-QCB interaction.  Such an interaction gives an opportunity
to probe QCB spectrum via IR absorption of the substrate (see below). 
Assuming the distance $D$ between the first array and the substrate to
be much smaller than the distance $d$ between arrays, one can keep
only the interaction between the substrate and the first array.  In
this case the system is described by the Hamiltonian
\begin{equation}
  H = H_0+H_{int}.
  \label{H}
\end{equation}
Its unperturbed part
\begin{eqnarray*}
    H_0=H_{K}+H_{1}+H_{2}
    \label{unperturbed}
\end{eqnarray*}
contains the Hamiltonians (\ref{arrays}) of the two arrays and the
kinetic energy of the substrate electrons with effective mass $m$ and
quadratic dispersion law $\varepsilon_{{\bf
k}}=\displaystyle{\frac{\hbar^{2}k^{2}}{2m}}$ (we omit the irrelevant
spin variables)
\begin{eqnarray*}
  H_{K} =
  \sum_{\bf{qm}}\varepsilon_{\bf{q+m}}c_{\bf{q+m}}^{\dag}c_{\bf{q+m}}.
  \label{Kinetic}
\end{eqnarray*}
Here as in the previous section, a quasi wave vector ${\bf q}$ belongs
to the first BZ and
${\bf{m}}$ is
a reciprocal lattice vector.
The interaction part of the Hamiltonian (\ref{H})
\begin{eqnarray*}
  H_{int}=H_C+H_{s1}+H_{12}.
  \label{H-int}
\end{eqnarray*}
includes inter-array interaction (\ref{inter}), Coulomb interaction
within the substrate
\begin{eqnarray*}
  H_C &=& \frac{1}{2}\sum_{\bf{qm}}U_{\bf{q+m}}
          \rho_{\bf{q+m}}^{\dag}\rho_{\bf{q+m}},\nonumber\\
  \rho_{\bf{k}}&=&\frac{1}{L}\sum_{\bf{k'}}c_{\bf{k'}}^{\dag}c_{\bf{k+k'}},
  \phantom{a}U_{\bf{k}}=\frac{2\pi e^2}{k},
  \phantom{a}{\bf{k}}={\bf{q}}+{\bf{m}}
  \label{Hc}
\end{eqnarray*}
and an interaction $H_{s1}$ between the substrate
and the first array. The latter interaction is a capacitive coupling
between charge fluctuations in the substrate and collective modes in the
nearest array. In coordinate representation it is written as
\begin{eqnarray*}
  H_{s1}&=&\sum_{n_2}\int dx_1dx'_1dx_2
  W(x_1-x'_1,x_2-n_2a)
  \times\nonumber\\&&\times
  \rho(x_1,x_2)\partial_{x'_1}\theta_1(x'_1,n_2a).
  \label{H-s1-coord}
\end{eqnarray*}
Here $\rho({\bf r})$ is the density operator for the substrate
electrons, $\sqrt{2}\partial_{x'_1}\theta_1(x'_1,n_2a)$ is the density
operator of the first array, and $W({\bf r}-{\bf r}')$ is Coulomb
interaction between the substrate and array
\begin{eqnarray*}
    W({\bf r})=
    \frac
    {\sqrt{2}e^2\zeta\left(
    \displaystyle{\frac{x_1}{r_0}}\right)}
    {\sqrt{\left|{\bf{r}}\right|^2+D^2}}.
\end{eqnarray*}
In momentum representation the interaction between the substrate and
the array has the form:
\begin{equation}
  H_{s1}= \sqrt{\frac{\hbar}{vg}}\sum_{\bf{mq}}W_{{\bf{q}}+{\bf{m}}}
             \rho_{\bf{q+m}}\theta_{1,{\bf{q+m}}_1}^{\dag},
  \label{Hs1}
\end{equation}
where
\begin{eqnarray*}
  W_{{\bf{k}}} = ik_1\sqrt{\frac{vg}{\hbar a}}\int dx_1dx_2
  W({\bf{r}})e^{i{\bf{kr}}}
\end{eqnarray*}
is proportional to the Fourier component of $W({\bf{r}})$.
\subsection{Dielectric Function}\label{subsec:Dielectric}
The high frequency properties of the system at zero temperature are
determined by the zeroes of the dielectric function
\begin{equation}
 \frac{1}{\epsilon({\bf{k}},\omega)}=
 1+U_{\bf{k}}\Pi({\bf k},\omega).
 \label{DielFun}
\end{equation}
Here
\begin{equation}
  \Pi({\bf{k}},\omega)=-\frac{i}{\hbar}\int\limits_{0}^{\infty}dt
  e^{i\omega t}
         \left\langle\left[
              \rho_{\bf{k}}(t),
              \rho_{\bf{k}}^{\dag}(0)
         \right]\right\rangle,
  \label{PolarCorrFun}
\end{equation}
is the polarization of the {\it substrate interacting with QCB},
$\rho_{\bf{k}}(t)=e^{iHt/\hbar}\rho_{\bf{k}}e^{-iHt/\hbar}$ is the
density of the {\it substrate} electrons in the Heisenberg
representation, and averaging is performed over the ground state of the
Hamiltonian (\ref{H}).\\

Dielectric properties of the substrate {\it per se} within the RPA
approach are described by the $2D$ version of the Lindhard formula
\begin{eqnarray}
 \epsilon_{s}({\bf{k}},\omega)&=&
 1-U_{\bf{k}}\Pi_{0}({\bf k},\omega),\nonumber\\
 \Pi_{0}({\bf k},\omega)&=&
    \frac{1}{L^2}\sum\limits_{\bf k'}
    \frac{\theta(\varepsilon_F-\varepsilon_{\bf k'})
    -\theta(\varepsilon_F-\varepsilon_{\bf k+k'})}
       {\hbar\omega-(\varepsilon_{\bf k+k'}-
                     \varepsilon_{\bf k'})+i0}.
 \label{Lindhard}
\end{eqnarray}
The {\it substrate} polarization $\Pi_{s}({\bf k},\omega)$ is
defined by the same Eq.(\ref{PolarCorrFun}) where the averaging in the
r.h.s. is performed with respect to the {\it substrate} Hamiltonian
$H_{s}=H_{K}+H_{C}.$ It can be written as
\begin{equation}
    \left(\Pi_{s}({\bf k},\omega)\right)^{-1}=
    \left(\Pi_{0}({\bf k},\omega)\right)^{-1}-U_{{\bf k}}
    \label{Pi_s},
\end{equation} 
that results in the expression for the dielectric function
\begin{equation}
 \frac{1}{\epsilon_{s}({\bf{k}},\omega)}=
 1+U_{\bf{k}}\Pi_{s}({\bf k},\omega)
 \label{DielFunSub}
\end{equation} 
similar to Eq.(\ref{DielFun}).\\ 

\begin{figure}[h]
\centerline{\epsfig{figure=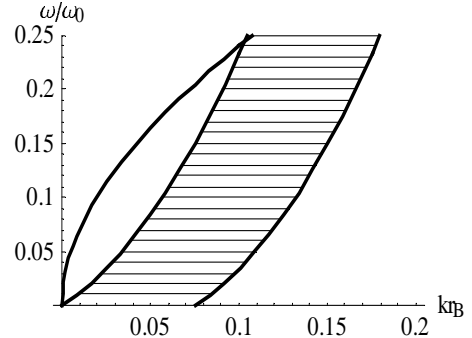,width=60mm,height=45mm,angle=0}}
\caption{Dispersion of the substrate plasmons (upper
line) and quasi-continuum spectrum of electron-hole excitations,
(dashed area). Frequency is measured in $\omega_0=v_F/r_B$ units.}
 \label{Disp0}
\end{figure}

The active branches of substrate excitations are the surface density
fluctuations which consist of $2D$ electron-hole pairs and surface
plasmon mode with dispersion law
\begin{equation}
 \omega_{s}(k)=v_F k\sqrt{1+\frac{1}{2 kr_B}},
 \label{SubstrDisp2}
\end{equation} 
where $r_B=\displaystyle{\frac{\hbar^2}{me^2}}$ (see e.g.
Ref.\onlinecite{Stern}).  The {R}{P}{A} spectrum of surface
excitations is shown in Fig.\ref{Disp0}.\\  

For GaAs, the substrate parameters are $m=0.068m_0$,
$m_0=9.1\cdot{10}^{-28}$~g is the mass of a free electron,
$v_F=8.2\cdot{10}^6$~cm/sec, $r_B=0.78$~nm,
$\omega_0=v_F/r_B={1.05}\times{10}^{14}$~sec$^{-1}$.  For
$k<k^{*}\approx 0.1r^{-1}_B$, the plasmon frequency lies above the
continuum spectrum of electron-hole pairs and the substrate plasmons
are stable.  Besides, one may easily satisfy the resonance condition
for the collective plasmon mode near the stability threshold $k\sim
k^{*}$ and QCB excitations with frequency $\sim
{10}^{14}$~sec$^{-1}$.  For large enough $k>k^{*}$ the plasmon
dispersion curve lies within the quasi-continuum spectrum and plasmons
become unstable with respect to decay into electron-hole pairs (Landau
damping of the substrate plasmons).\\

To obtain the dielectric function of the whole system substrate plus
QCB (\ref{DielFun}), one should take into account that Umklapp
processes stimulated by interaction between the substrate and the
first array (\ref{Hs1}) as well as the interaction between arrays
(\ref{inter}), produce modes with wave vectors ${\bf{q}}+{\bf{m}}$
with various reciprocal lattice vectors ${\bf{m}}$.  This necessarily
leads to appearance of non-diagonal components of the polarization
operator
$$
  \Pi({\bf q} +{\bf m},{\bf q} +{\bf m}';\omega)=
  -\frac{i}{\hbar}\int\limits_{0}^{\infty}dt
  e^{i\omega t}
  \left\langle\left[
             \rho_{\bf{q+m}}(t)
             \rho_{\bf{q+m'}}^{\dag}(0)
  \right]\right\rangle.
$$
In what follows we always consider a fixed frequency $\omega$ and a
fixed wave vector ${\bf q}$ in the BZ. Henceforth the variables ${\bf
q}$ and $\omega$ are omitted below for simplicity.  In the framework
of {R}{P}{A} approach, $\Pi({\bf{m}},{\bf{m}}')=$
{\epsfig{figure=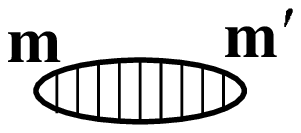,width=10mm,height=5mm,angle=0}} satisfies
the following Dyson-type equation
\begin{equation}
  {\mbox{\epsfig{figure=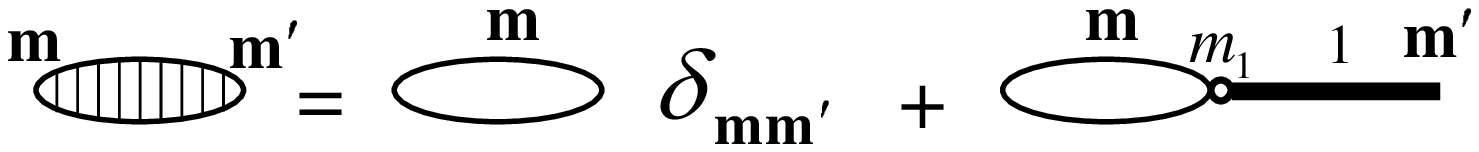,width=70mm,height=6mm,angle=0}}}.
  \label{EqDyson1}
\end{equation}
Here the first term 
${\mbox{\epsfig{figure=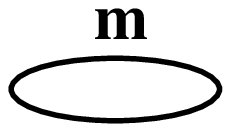,width=10mm,height=5mm,angle=0}}}=
\Pi_s({\bf {m}})$ 
in the right hand side is the substrate polarization (\ref{Pi_s}) of the 
isolated substrate itself, the empty point 
{\epsfig{figure=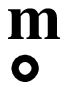,width=2.2mm,height=3.6mm,angle=0}} 
$={W}_{\bf{m}}\equiv{W}_{{\bf{q}}+{\bf{m}}}$ is a simple vertex which   
describes substrate - (first) array interaction, and the thick line
\begin{eqnarray}
  {\epsfig{figure=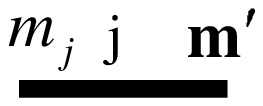,width=10mm,height=4mm,angle=0}}=
-\frac{i}{\hbar}\int\limits_{0}^{\infty}dt
  e^{i\omega t}
  \left\langle\left[
       \theta_{j{\bf{q+m}}_j}(t),
       \rho_{{\bf{q}}+{\bf{m}}'}^{\dag}(0)
  \right]\right\rangle
  \label{Dj}
\end{eqnarray}
is the correlation function of the $j$-{t}{h} array mode and the
substrate plasmon.\\

The Dyson equation (\ref{EqDyson1}) should be completed by two equations
for the correlation functions (\ref{Dj})
($j=1,2$)
\begin{eqnarray}
  &&
  {\mbox{\epsfig{figure=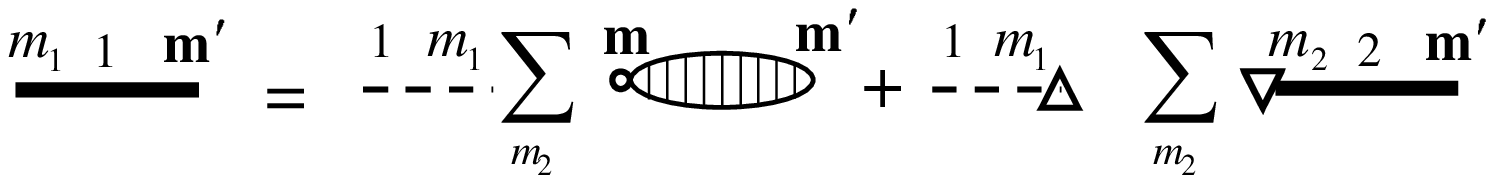,width=70mm,height=7mm,angle=0}}}
  \label{EqDyson2}
  \\
  &&
  {\mbox{\epsfig{figure=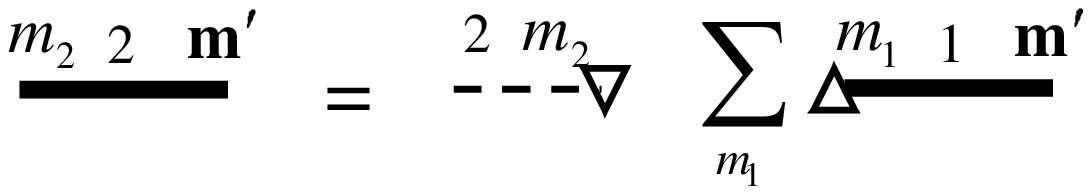,width=43mm,height=7mm,angle=0}}}.
  \label{EqDyson3}
\end{eqnarray}
Here the dashed line
{\mbox{\epsfig{figure=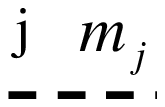,width=10mm,height=5mm,angle=0}}}
$=D_{j}^{0}(m_j)$ ($j=1,2$) is the bare correlation function of
the $j$-{t}{h} array modes
\begin{eqnarray*}
  D_{j}^{0}(m_j) &=&
  -\frac{i}{v g}\int\limits_{0}^{\infty}dt
  e^{i\omega t}\left\langle\left[
                    \theta_{j{\bf{q+m}}_j}(t),
                    \theta_{j{\bf{q+m}}_j}^{\dag}(0)
               \right]\right\rangle_{0}
  \nonumber\\&=&
  \frac{1}{\omega^2-v^2(q_j+m_j)^2},
  \label{D-j0}
\end{eqnarray*}
and two triangles 
{\epsfig{figure=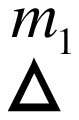,width=3mm,height=5mm,angle=0}} 
$=\sqrt{\hbar\phi/vg}{\xi}_{q_{1}+m_{1}Q}$ and 
{\epsfig{figure=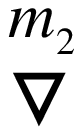,width=3mm,height=5mm,angle=0}} 
$=\sqrt{\hbar\phi/vg}{\xi}_{q_{2}+m_{2}Q}$ form the second vertex 
describing the separable inter-array interaction (\ref{Explan}).\\

Solving the system of equations (\ref{EqDyson1}), (\ref{EqDyson2}) and
(\ref{EqDyson3}) one obtaines the diagonal element 
$\Pi({\bf m})\equiv \Pi({\bf m,m})$ of the polarization operator
\begin{equation}
  \left[\Pi({\bf m})\right]^{-1}=
  \left[\Pi_s({\bf m})\right]^{-1}-
        |W_{\bf{m}}|^2D({\bf m}).
  \label{Polar-Solution}
\end{equation}
The second term in the right-hand side of this equation describes 
renormalization of this vertex by interaction between the substrate 
and QCB. The factor $D({\bf m})$ is a renormalized correlation 
function of modes of the first array
\begin{eqnarray}
  \left[
D({\bf m})
\right]^{-1}&=&
  \left[
  D_1^0(m_1)
  \right]^{-1}
  -(w({\bf m})+\varphi(m_1)).
  \label{Dm}
\end{eqnarray}

Here $w({\bf{m}})$ describes the effective interaction 
between the first array and substrate
\begin{eqnarray}
  w({\bf{m}}) &=&F(m_{1})-
  |W_{\bf{m}}|^2\Pi_s({\bf m}),\nonumber\\
  F(m_{1})&=&\sum_{m_{2}}|W_{\bf{m}}|^2\Pi_s({\bf m})
	   \label{w}
\end{eqnarray}
and $\varphi(m_1)$ is the effective interaction between the two arrays
\begin{eqnarray}
  \frac{\omega_{m_{1}}^{2}}{\varphi(m_1)} &=&
  \left[
  \phi^{2}\sum\limits_{m_2}\omega_{m_{2}}^{2}D_2^0(m_2)
  \right]^{-1}-\Psi_{m_{1}},\\
  \omega_{m_{j}}&=&v(q_{j}+m_{j}Q),\nonumber
  \label{varphi}
\end{eqnarray}
renormalized by Coulomb interaction of array modes with substrate 
plasmons, 
\begin{eqnarray}
   \Psi_{m_1} &=&
         \sum_{{m'_1}\ne{m_1}}
         \frac{\omega_{m'_{1}}^{2}}
         {\left(D_1^0(m'_1)
         \right)^{-1}-F(m'_1)}.
  \label{M}
\end{eqnarray}
Equations (\ref{Polar-Solution}) - (\ref{M}) together with the 
definition (\ref{DielFun}) solve the problem of dielectric properties 
of the combined system QCB-substrate within RPA.\\

The whole $\{\omega,{\bf k}\}$ space can be divided into two parts. 
Within the first subspace the dielectric function (\ref{DielFun}) is a
real function.  Its zeros define the spectrum of collective modes. 
Within the second subspace the dielectric function has non-zero
imaginary part.  The latter is related to regions where these
excitations are unstable.  These two subspaces will be considered
separately in the next two subsections.
\subsection{Excitations}\label{subsec:Renorm}
The spectrum of collective excitations in QCB-substrate system is
determined by zeroes of the dielectric function
$\epsilon({\bf{q}},\omega)=0.$ In the low frequency limit
$\omega\ll{v}{Q}/2$ taking into account only contribution of the
lowest QCB band (i.e. omitting the terms with ${\bf{m'}}\neq 0$), we
obtain the dispersion equation
\begin{eqnarray}
  &&\left(\omega^2-\omega_{s}^2(q)\right)
  \bigg[
       \left(\omega^2-(1-\alpha)v^2q_1^2\right)
       \left(\omega^2-(1-\alpha)v^2q_2^2\right)-
       \nonumber\\
	   &&-\phi^{2}
       \xi^2_{q_1}\xi^2_{q_2}
  \bigg]-
  \frac{2m}
  {\pi^2\hbar^2}
  v_F^2q^2|W_{\bf{q}}|^2
  \left(\omega^2-v^2q_2^2\right) = 0,
  \label{Eq-omega-2}\\
      &&\phantom{aaaaaaaaaaaaaaaaa}
	  \alpha=\phi^{2}\frac{a}{r_{0}},\nonumber
\end{eqnarray}
where $\omega_{s}(q)$ is the substrate plasmon frequency
(\ref{SubstrDisp2}).  Neglecting both inter-array ($\propto\phi^{2}$)
and substrate-first array ($\propto W^{2}_{{\bf{q}}}$) interactions,
we are left with the substrate plasmon mode and the modes of two
arrays.  Asymmetry of the dispersion relation (\ref{Eq-omega-2}) with
respect to the two QCB arrays reflects the fact that the substrate
interacts only with the first array.\\

The key question is the robustness of the QCB spectrum against
interaction with $2D$ substrate excitations.  To answer this question
we note that for wave vector ${\bf q}\rightarrow 0$ both interactions
vanish
\begin{equation}
    \xi_{q_{1}}^{2}\xi_{q_{2}}^{2},|W_{\bf{q}}|^2\rightarrow0
    \label{longwave}
\end{equation}
for all directions of ${\bf q}$ except the first array direction
$x_{1}$ (the latter direction is singular in all similar $1D-2D$
problems).  Therefore in the long wave limit $q\ll Q$ the interaction
just renormalizes the bare dispersion laws of the arrays
\begin{eqnarray*}
    \frac{\omega_{j}^{2}}{v^{2}q_{j}^{2}}=
    1-\alpha-\delta_{1j}\alpha_{{\bf{q}}},\phantom{aa}
	\alpha_{{\bf{q}}}=
    \frac{1}{2\pi v^{2}e^{2}}
    \frac{q|W_{\bf{q}}|^2}{q_{1}^{2}},
    \label{LL}
\end{eqnarray*}
conserving its LL linearity. This result verifies stability of QCB
plasmons with respect to substrate-QCB interaction.\\

\begin{figure}[h]
\centerline{\epsfig{figure=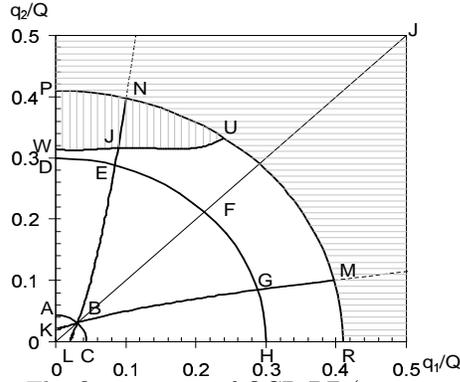,width=60mm,height=50mm,angle=0}}
\caption{The first quarter of QCB BZ ($a=20$~nm). QCB plasmons are in resonance 
with surface plasmons along the lines $OM$ and $ON.$ The surface 
plasmons are stable within the internal part of BZ bounded 
by the curve $WJU$ and the arc $UMR.$ }
\label{BZ-S-QB}
\end{figure}

The dispersion relation (\ref{Eq-omega-2}) shows that the
QCB-substrate interaction results in appearance of two more resonant
directions in the BZ. In addition to the diagonal line $OJ$
(Fig.\ref{BZ-S-QB}) that is the resonance line for inter-array
interaction, two new lines arise, which correspond to resonant
interaction of the substrate with the first or the second array.  The
resonance condition $\omega_{s}(q)=vq_j\equiv\omega_j({\bf{q}})$ is
fulfilled along the line $LN$ for $j=1$ and along the line $KM$ for
$j=2.$ The points $J,M$ on these lines are the points of onset of
Landau damping.  The region of Landau damping exists outside the curve
$WJUMR$ and consists of two subregions.  One of them exists in an
isolated substrate outside the arc $PNUMR.$ Another region is secluded
within the close curve $WJUNPW$ and appears exclusively due to
QCB-substrate interaction (see subsection \ref{subsec:Damping} for
details).\\

Far from the symmetric point $B$, the two renormalized dispersion 
laws along $LN$ are
\begin{equation}
 \omega_{\pm}^2=v^2q_{1}^{2}
 \left(
      1\pm
      \sqrt{\alpha_{\bf{q}}}
 \right).
 \label{Res-Sub-Arr1}
\end{equation}
They describe the splitting of two coupled substrate-first array
modes.  The third mode is slightly renormalized plasmon propagating
along the second array.  Along the resonance line $KM$ one of the
dispersion laws conserves its bare form $$\omega=vq_{2}$$ while
another one transforms into
\begin{equation}
      \frac{\omega^2}{ v^2q_{2}^{2}} =
      1-(\alpha_{\bf{q}}+\alpha)
      \frac{q_1^2}{q_1^2-q_2^2}.
\label{Res-Sub-Arr2}
\end{equation}
These two modes are coupled substrate plus second array modes.  The
third mode represents slightly renormalized plasmon propagating along
the first array.  Along the diagonal $OJ$, the two array modes are in
resonance $vq_1=vq_2\equiv\omega({\bf{q}})$.  Here we get two coupled
QCB plasmons with dispersion laws
\begin{equation}
  \omega_{\pm}=vq
  (1\pm\sqrt{\alpha})
  \label{Res-Arr1-Arr2}
\end{equation}
and substrate plasmon with slightly renormalized frequency.  At the
symmetric point $B,$ $q_{1B}=q_{2B}\approx{0.1Q},$ all three modes are
in resonance $\omega_{s}(q_B)=vq_{1B}=vq_{2B}\equiv\omega_B$. 
Interaction lifts the degeneracy and renormalizes the frequencies
\begin{equation}
 \omega_0^2=\omega_B^2, \ \ \ \
 \omega_{\pm}^2=\omega_B^2
 \left(1\pm\sqrt{\alpha_{{\bf{q}}_B}+\alpha}\right).
 \label{Res-symm-point}
\end{equation}

\begin{figure}[h]
\centerline{\epsfig{figure=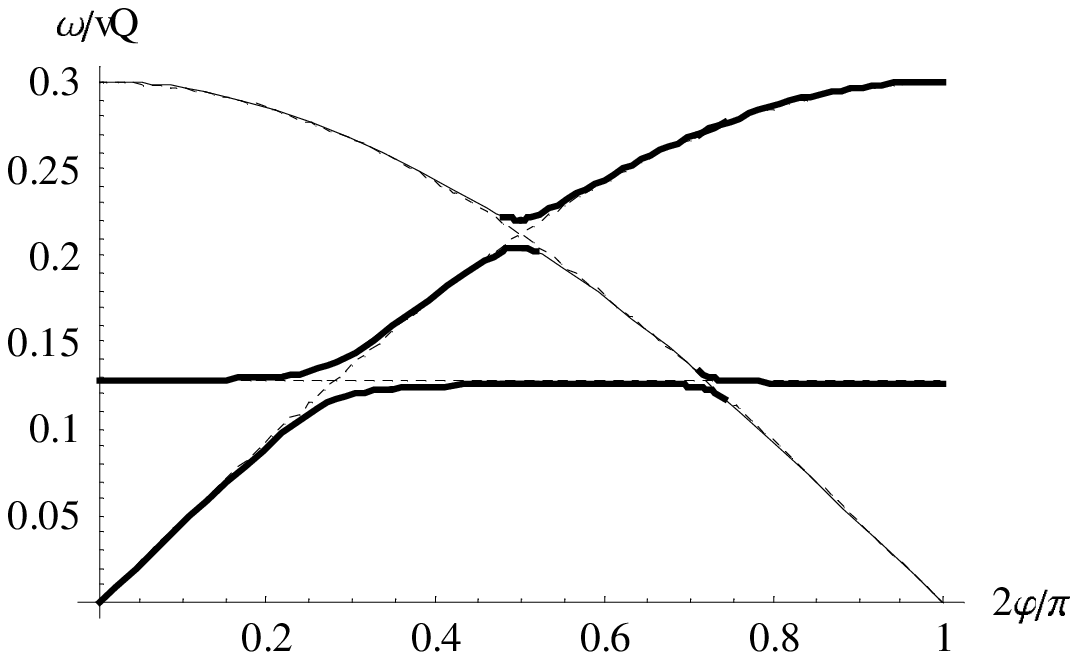,width=60mm,height=50mm,angle=0}
}
\caption{Frequencies of substrate plasmons and modes of {Q}{C}{B}
with interaction between them (solid lines) and without the
interaction (dashed lines) as functions of the azimuthal angle $\varphi$
for $q=0.3 Q$.}
\label{Sp-off}
\end{figure}

\begin{figure}[h]
\centerline
{\epsfig{figure=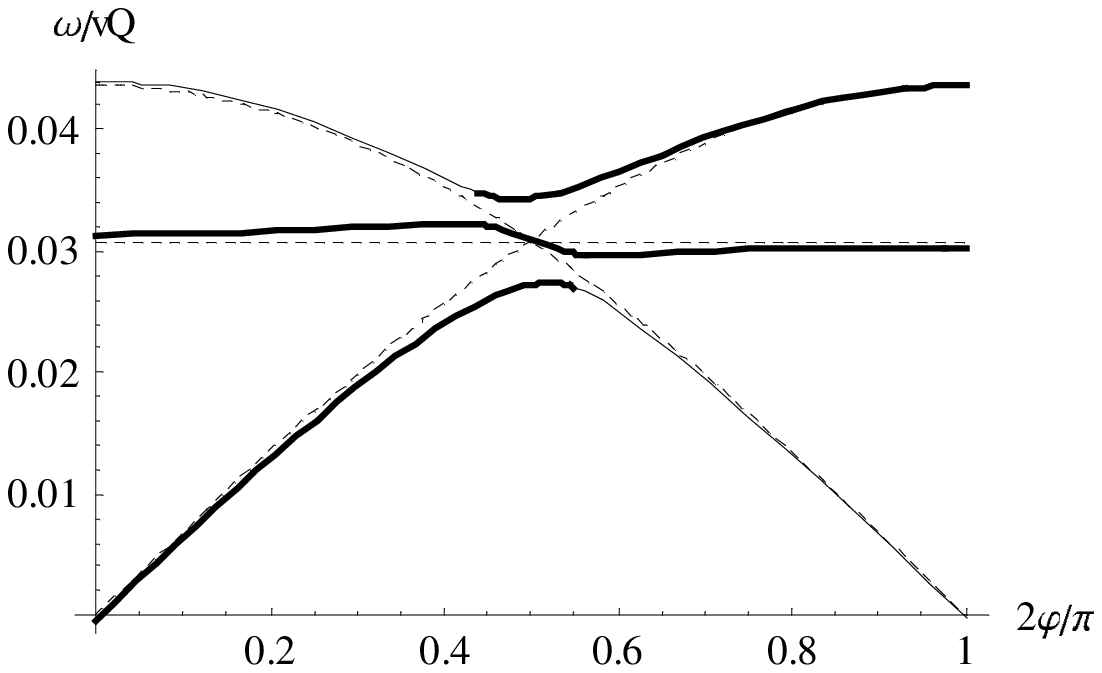,width=60mm,height=50mm,angle=0}}
\caption{Frequencies of substrate plasmons and modes of {Q}{C}{B}
with interaction (solid lines) and without interaction (dashed
lines)  as functions of azimuthal angle $\varphi$ for
$q=0.03 Q$.}\label{Sp-on}
\end{figure}

In Figs.  \ref{Sp-off}, \ref{Sp-on} the dispersion curves calculated
along the circle arcs with two different radii are shown.  Solid lines
show the dispersion curves of interacting system whereas the dashed
lines describe those of noninteracting substrate and QCB. Here the 
azimuthal angle $\varphi$ is introduced as in the previous section.\\

We start with a general case (Fig.\ref{Sp-off}) that represents the
dispersion curves along the $DEFGH$ line for $q=0.3Q.$ Points
$D,E,F,G,H$ in Fig.\ref{BZ-S-QB} correspond to values
${2\varphi}/{\pi}=0,\phantom{a}0.28,\phantom{a}0.5,
\phantom{a}0.72,\phantom{a}1.$ Far from the resonance lines (i.e. not
too close to the points $E$, $F$ and $G$) the dispersion curves
describe slightly renormalized frequencies of the modes propagating
within the substrate and along two arrays.  At the point $E$ in Fig. 
\ref{BZ-S-QB}, the modes propagating in the substrate and along the
first array are degenerate.  The substrate-QCB interaction splits this
degeneracy and the renormalized frequencies are described by
Eq.(\ref{Res-Sub-Arr1}).  At the point $F$, the modes propagating
along the first and second arrays are degenerate and the frequencies
are described by Eq.(\ref{Res-Arr1-Arr2}).  At the point $G$, the
modes propagating in the substrate and along the second array are
degenerate and the frequencies are described by
Eq.(\ref{Res-Sub-Arr2}).  For any fixed angle $\varphi$ three
absorption lines should exist, but the intensity of that related to
the second array is proportional to both (weak) substrate-array and
inter-array interaction, so this line is practically not observable. 
The corresponding dispersion curve is plotted as a thin solid line. 
On the other hand, the thick solid lines are used for dispersion
curves corresponding to the substrate plasmons and excitations of the
first array.\\

Dispersion curves along the line $ABC,$
($\frac{2\varphi}{\pi}=0,\phantom{a}0.5,\phantom{a}1.0$) that includes
the triple resonance point $B,$ are displayed in Fig.\ref{Sp-on} (here
$q\approx0.03 Q$).  Here at point $B$, all three modes are strongly
coupled and the frequencies are described by
Eq.(\ref{Res-symm-point}).  As in the previous figure, dispersion
curve corresponding to the second array plasmons is plotted by a thin
solid line.\\
\subsection{Landau Damping}\label{subsec:Damping}
Dielectric losses of an isolated substrate are described by an
imaginary part of its dielectric function $\epsilon_{s}$
(\ref{DielFunSub}).  This imaginary part is non-zero within the dashed
region in Fig.\ref{Disp0} due to appearance of imaginary part of the
bare polarization operator $\Pi_0({\bf k},\omega)$ (\ref{Lindhard})
\begin{eqnarray*}
  \Im\Pi_0({\bf k},\omega) &&=
  \frac{m}{2\pi\hbar^2}
  \frac{1}{\kappa^2}
  \left[
      \theta(\kappa^2-\nu_{+}^2)
       \sqrt{\kappa^2-\nu_{+}^2}-
  \right.\nonumber\\&&\left.-
      \theta(\kappa^2-\nu_{-}^2)
       \sqrt{\kappa^2-\nu_{-}^2}
  \right],\nonumber\\
  \kappa &&=\frac{k}{k_{F}},\ \nu_{\pm}=\nu\pm\frac{\kappa^{2}}{2},
  \ \nu=\frac{\omega}{v_{F}k_{F}}.
 \label{ImPolarOper0}
\end{eqnarray*}
  As was mentioned above, the plasmon dielectric losses
are related to the Landau damping of collective excitation 
with momentum $k>k^{*},\ \ k^{*}r_{B}\approx 0.1$ into
electron-hole pair.\\

The substrate-QCB interaction remarkablly changes the conventional
picture of substrate plasmon dielectric losses.  The imaginary part of
the dielectric function $\epsilon({\bf k},\omega)$ (\ref{DielFun}) of
the system differs from zero not only within the dashed region.  New
regions of Landau damping appear due to QCB - substrate interaction. 
Indeed, QCB serves as a diffraction grid for the substrate like in the
case considered in Section \ref{sec:Absorption}.  The corresponding
Umklapp processes initiate Landau damping of surface plasmons within
the long wave region $k< k^{*}$ where the usual Landau damping is
absent.  The main contribution is related to the renormalization term
$w({\bf m})$ in Eq.(\ref{w}) due to Umklapp processes along the
$x_{2}$ axis (summation over $m_{2}$ in the expression for function
$F(m_{1})$ is implied).  It is proportional to the fourth power of QCB
- substrate interaction $W^{4}$.  The Umklapp processes along both
directions $x_{1,2}$ contribute also to the renormalization term
$\varphi(m_{1})$ in Eq.(\ref{Dm}).  However, they contain additional
small parameter $\phi^{4}$ related to inter-array interaction within
QCB and we do not take them into account.\\

\begin{figure}[h]
\centerline{\epsfig{figure=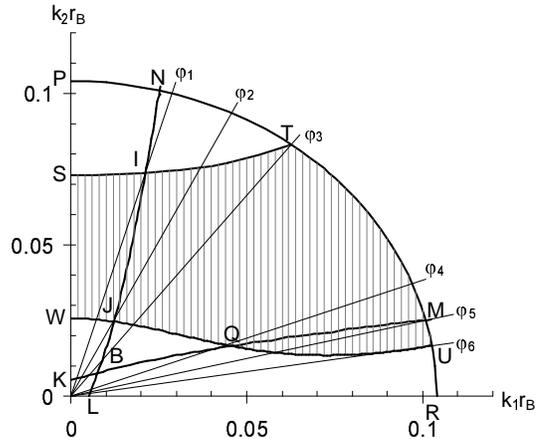,width=70mm,height=58mm,angle=0}}
\caption{``Phase diagram'' describing possible types of new regions of 
Landau damping. Superlattice constant is chosen as $a=30$~nm. For an 
isolated substrate Landau damping occurs outside the arc $PNTMUR.$ New 
regions are bounded by curves $SIT$ and $WJQU.$}
\label{area}
\end{figure}

These new Landau damping regions can be described in detail with the
help of a ``phase diagram'' displayed in Fig.\ref{area}.  Coordinate
axes in the figure correspond to the surface plasmon momenta $k_{1,2}$
along $x_{1,2}$ directions measured in $r_{B}^{-1}$ units.  The QCB
period is $a=30$~nm.  Such a choice enables us to realize a rich
variety of possible damping scenarios.\\

The Landau damping region of the isolated substrate, defined by
$|k|r_{B}>0.1$, is situated outside the circular arc $PNTMUR.$ A new
damping region inside this arc is bounded by two curves $SIT$ and
$WJQU.$ The line $LBJIN$ ($KBQM$) corresponds to resonance between
substrate plasmon and the first (second) array QCB plasmon.  The
points $I$, $J$, $T$, $Q$, $M$, $U$ of intersection between the
resonant lines, the curves $SIT$ and $WJQU$, and the arc $PNTMUR$
define six rays $OI$, $OJ$, $OT$, $OQ$, $OM$, and $OU$ and
corresponding six angles $\varphi_{1}=16^{\circ},$
$\varphi_{2}=33^{\circ},$ $\varphi_{3}=38^{\circ},$
$\varphi_{4}=69^{\circ},$ $\varphi_{5}=76^{\circ},$
$\varphi_{6}=81^{\circ}.$ Each pair of adjacent rays (together with
coordinate semiaxes) bounds a specific structure of new damping
region.\\

New damping region corresponding to the substrate plasmon propagating
within the first sector $\varphi<\varphi_{1}$ is separated quite well
from the region of initial Landau damping.  In fact it presents a new
damping band.  Its boundaries are defined by inersection point of the
ray with a fixed angle $\varphi$ with the curves $SIT$ and $WJQU.$ The
damping amplitude is small because of the small factor of order
$W^{4}$ mentioned above.  When the ray tends to the $\varphi_{1}$
direction, small peak appears near the ``blue'' boundary of the new
damping region (precursor of the resonance between the substrate
plasmon and the first array QCB plasmon).\\

\begin{figure}[h]
\centerline{\epsfig{figure=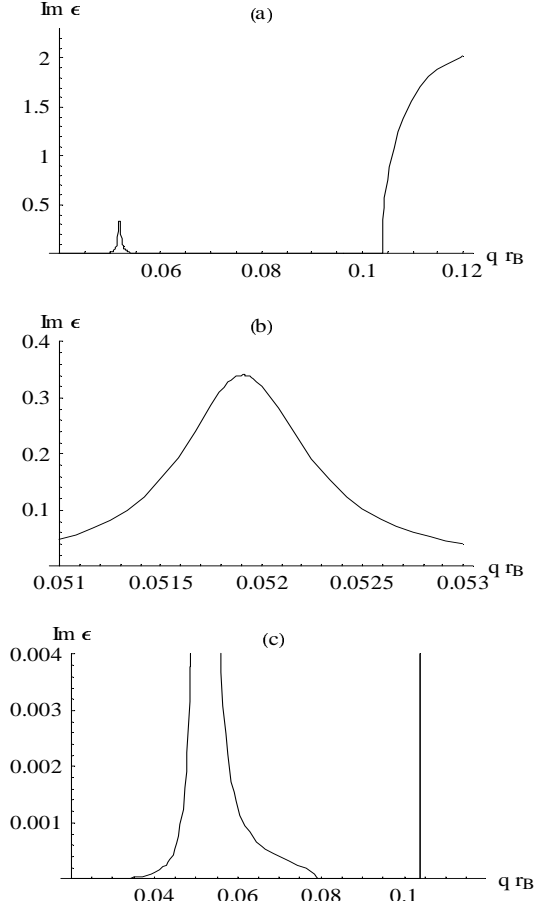,width=70mm,height=120mm,angle=0}}
\caption{Additional damping region for $\varphi=20^{\circ}.$ {\it a)}
main peak and initial Landau damping region; {\it b)} central part of
the peak zoomed; {\it c)} total new damping region and edge of the
initial damping region}
\label{Res}
\end{figure}

Within the second sector $\varphi_{1}<\varphi<\varphi_{2},$ well
pronounced resonant peak appears within the damping band.  It is
described in detail in Fig.\ref{Res} for $\varphi=20^{\circ}.$ The
peak amplitude is of the order of damping amplitude within the initial
damping region (panel {\it a}).  It has a well pronounced Lorentzian
form (panel {\it b}) placed on the wide and low "pedestal" (panel {\it
c}).  \\

\begin{figure}[h]
\centerline{\epsfig{figure=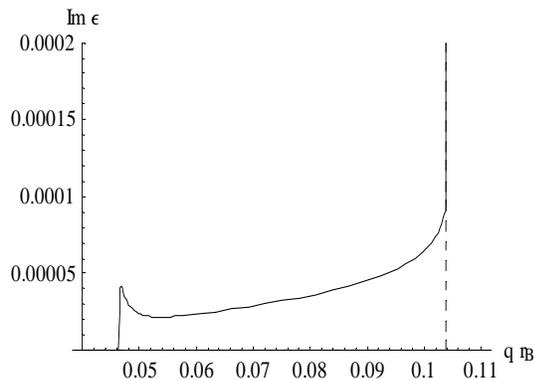,width=70mm,height=50mm,angle=0}}
\caption{Damping tail for $\varphi=68^{\circ}$ and the 
initial damping region. Precursor of the resonant peak is resolved 
quite well.}
\label{NonRes}
\end{figure}

The next two sectors $\varphi_{2}<\varphi<\varphi_{3}$ and
$\varphi_{3}<\varphi<\varphi_{4},$ do not contain resonant peaks at
all.  The new damping region corresponding to the first one of them is
still separated from the initial damping region touching it at only at
the angle $\varphi_{3}.$ For $\varphi_{3}<\varphi<\varphi_{4},$ the
new damping band is altered by a weak damping tail.  The damping
amplitude in the tail is small due to the same reasons mentioned
above.  It is displayed in the insertion to the Fig.\ref{NonRes} for
the angle $\varphi=68^{\circ}.$ This angle is close to the sector
boundary and a precursor of the resonant peak is well pronounced.\\

Further increase of the angle $\varphi_{4}<\varphi<\varphi_{5}$ leads
to re-appearance of the resonant peak within the tail.  In this case
one deals with a resonance between the substrate plasmon and the QCB
plasmon in the second array.  Existence of this resonance is caused by
inter-array interaction that brings additional small parameter to the
imaginary part of the dielectric function.  As a result, the width of
the peak is extremely smaller in the second sector while surprisingly
the peak amplitude has the same order of magnitude as in the case of
resonance with the nearest to substrate (first) array.\\

Within the sixth sector $\varphi_{5}<\varphi<\varphi_{6}$ the damping
tail does not contains any resonance.  It vanishes at
$\varphi=\varphi_{6}$ and within the last seventh sector
$\varphi_{6}<\varphi<\pi/2$ the damping occurs only within the initial
landau damping band.\\

The existence of the additional QCB band (tail) of Landau
damping and appearance of the resonant peak within the band (tail) is
a bright manifestation of interplay between really $2D$ surface
plasmons and quasi-$2D$ QCB plasmons.\\
\section{Conclusion}\label{sec:Conclu}
In conclusion, we investigated the possibility of spectroscopic
studies of the excitation spectrum of quantum crossbars, which
posesses unique property of dimensional crossover both in spatial
coordinates and in $({\bf q},\omega)$ coordinates.  It follows from
our studies that the plasmon excitations in QCB may be involved in
resonance diffraction of incident electromagnetic waves and in optical
absorption in the IR part of spectrum.\\  

In the case of direct interaction of external electric field with QCB,
infrared absorption strongly depends on the direction of the wave
vector ${\bf q}$.  One can observe dimensional crossover from $1D\to
2D$ behavior of QCB by scanning an incident angle.  The crossover
manifests itself in the appearance of a set of absorption doublets
instead of the set of single lines.  At special directions, one can
observe new type of crossover where doublets replace the single lines
with changing frequency at a fixed ${\bf q}$ direction.\\

Capacitive contact between QCB and semiconductor substrate does not 
destroy the LL character of the long wave excitations. However, 
quite unexpectedly the 
interaction between the surface plasmons and plasmon-like excitations 
of QCB essentially influences the dielectric properties of a substrate. 
First, combined resonances manifest themselves in a complicated 
absorption spectra. Second, the QCB may be treated as the difraction 
grid for a substrate surface, and Umklapp diffraction processes 
radically change the plasmon dielectric losses. So the surface plasmons
are more fragile against interaction with superlattice of quantum wires
than the LL plasmons against interaction with 2D electron gas in a substrate. \\

Dimensional crossover in QCB plays a significant role in all the above
phenomena.\\
\section*{Acknowledgments}

S.G. appreciates discussions with N. Beletskii concerning excitation 
of surface plasmon in semiconducting substrate. 
S.G. and I.K. are thankful to V. Liubin and M. Klebanov for discussions of 
various ways of experimental observation of optical absorption in 
thin films. This research is partially supported by grants from 
Israeli Science Foundation and US-Israel Binational Science Foundation. 

\end{document}